\documentclass[12pt,preprint]{aastex}

\shortauthors{Matthews et al.}
\shorttitle{The \HI\ Tail of Mira}

\begin{document}

\newcommand{\ang}{\rm \AA}
\newcommand{\msun}{M$_\odot$}
\newcommand{\lsun}{L$_\odot$}
\newcommand{\days}{$d$}
\newcommand{\degree}{$^\circ$}
\newcommand{\ud}{{\rm d}}
\newcommand{\as}[2]{$#1''\,\hspace{-1.7mm}.\hspace{.0mm}#2$}
\newcommand{\am}[2]{$#1'\,\hspace{-1.7mm}.\hspace{.0mm}#2$}
\newcommand{\ad}[2]{$#1^{\circ}\,\hspace{-1.7mm}.\hspace{.0mm}#2$}
\newcommand{\lsim}{~\rlap{$<$}{\lower 1.0ex\hbox{$\sim$}}}
\newcommand{\gsim}{~\rlap{$>$}{\lower 1.0ex\hbox{$\sim$}}}
\newcommand{\HA}{H$\alpha$}
\newcommand{\HII}{\mbox{H\,{\sc ii}}}
\newcommand{\kms}{\mbox{km s$^{-1}$}}
\newcommand{\HI}{\mbox{H\,{\sc i}}}
\newcommand{\KI}{\mbox{K\,{\sc i}}}
\newcommand{\nan}{Nan\c{c}ay}
\newcommand{\galex}{{\it GALEX}}
\newcommand{\jks}{Jy~km~s$^{-1}$}

\title{Discovery of an \HI\ Counterpart to the Extended Tail of Mira}
\author{L. D. Matthews\altaffilmark{1}, Y. Libert\altaffilmark{2},
E. G\'erard\altaffilmark{3}, T. Le~Bertre\altaffilmark{2}, 
M. J. Reid\altaffilmark{1}}

\altaffiltext{1}{Harvard-Smithsonian Center for Astrophysics,
60 Garden Street, Cambridge, MA, USA 02138}
\altaffiltext{2}{LERMA, UMR 8112, Observatoire de Paris, 61 av.
de l'Observatoire, F-75014 Paris, France}
\altaffiltext{3}{GEPI, UMR8111, Observatoire de Paris, 5 Place J.
Janssen, F-92195 Meudon Cedex, France}

\begin{abstract}
We report the detection of an \HI\ counterpart to the
extended, far-ultraviolet-emitting
tail associated with the 
asymptotic giant branch star Mira ($o$~Ceti). Using the 
\nan\ Radio Telescope (NRT), we have detected
emission as far as $88'$ north of
the star, confirming that the tail contains a significant atomic component
($M_{\rm HI}\sim4\times10^{-3}~M_{\odot}$). 
The NRT spectra reveal a deceleration of the tail gas caused by
interaction with the local interstellar medium. We estimate an age 
for the tail of $\sim1.2\times10^{5}$~years,
suggesting that the mass-loss history of Mira has been more prolonged than
previous observational estimates. Using the
Very Large Array (VLA) we have also imaged the \HI\ 
tail out to $\sim12'$ (0.4~pc)
from the star. The detected emission shows a ``head-tail'' morphology, but
with complex substructure. Regions with detected \HI\ emission
correlate with far-ultraviolet-luminous regions on large scales, 
but the two tracers are not 
closely correlated on smaller scales ($\lsim1'$). 
We propose that detectable tails
of \HI\ are likely to be a 
common feature of red giants undergoing mass-loss.
\end{abstract}


\section{Introduction}
Mira ($o$~Ceti) is a mass-losing star on the asymptotic giant
branch (AGB). 
It is the archetype of a class of pulsating, long-period
variables, characterized by regular pulsations (with periods
of order hundreds of days) and large-amplitude variations in optical brightness
(by up to $\sim$8 mag; e.g., Reid \& Goldston 2002).
Mira is also a member of a wind-accreting
binary system, Mira~AB, with a projected separation of $\sim$\as{0}{5} 
($\sim$54~AU; Matthews \& Karovska
2006)\footnote[4]{All physical quantities
quoted in this paper assume a distance of 107~pc (Knapp et al. 2003).}.

Despite being the subject of observational scrutiny for many decades,
Mira continues to yield surprises. Recently, using
far-ultraviolet (FUV) imaging data from the
\galex\ satellite, Martin et
al. (2007; hereafter M07) discovered that Mira is surrounded 
by a bow shock structure
and sports a spectacular cometary-like tail,
stretching two degrees ($\sim$4~pc) on the sky. 
Mira has a rather high space velocity [$\sim$128~\kms\ with respect to the
interstellar medium (ISM); see \S~\ref{NRT}], and the tail extends 
backwards along 
its direction of motion. This tail is believed to arise from the 
interaction of Mira's wind with the ambient medium as 
the star moves supersonically through 
the ISM. The tail
is the first of its kind ever discovered, and M07 proposed that its FUV
emission  arises from H$_{2}$ molecules that are
collisionally excited by turbulent mixing between the cool molecular gas and
the electrons from a shock-heated gas component. 

We recently observed Mira in the \HI\ 21-cm line using the \nan\ Radio
Telescope (NRT) and the Very Large
Array (VLA)\footnote[5]{The Very Large Array of the National Radio Astronomy
Observatory is a facility of the
National Science Foundation, operated under cooperative agreement by
Associated Universities, Inc.} as part of a larger, ongoing 
\HI\ survey program of the
circumstellar envelopes of evolved stars (see G\'erard \& Le~Bertre 2006; 
Matthews \& Reid 2007). \HI\ is now known to be common in
circumstellar environments
and frequently shows evidence of extending to very large distances
from the star ($\gsim$1~pc; 
G\'erard \& Le~Bertre 2006 and references therein). 

Mira was previously observed in \HI\ by Bowers \& Knapp (1988), and for
more than a decade remained the only AGB star known to have associated \HI\
emission. 
Unfortunately, the data of Bowers \& Knapp had a
signal-to-noise too low to permit a detailed investigation of the morphology
and kinematics of the circumstellar material, although 
these authors did report
tentative evidence for an interaction between the circumstellar debris
and the surrounding ISM. 
More recently, NRT observations of Mira by G\'erard \& Le~Bertre (2006) 
provided a significantly improved 
\HI\ line profile showing a roughly triangular shape, similar to that
previously seen in CO (e.g., Winters et al. 2003), as well as evidence
for a northward extension of the emission. 
However, the  coarse spatial resolution of the \nan\ beam
provided little detail on the
\HI\ distribution close to the star. To better characterize the extent and
morphology of the \HI\ envelope of Mira, 
we therefore obtained new imaging observations
with the VLA. As we describe here, the fortuitous timing
of our observations 
provides a powerful complement to the recent {\it GALEX} results for
understanding the mass-loss history of Mira. 
To probe the most extended,
lowest column density material in the recently discovered FUV tail, 
we have also obtained new mapping observations of an extended region 
around Mira using the NRT.

\section{VLA Observations\protect\label{VLA}}
Mira was observed in the \HI\ 21-cm line with the VLA on 2007 April
1, April 30 and  May 11 using the most compact (D)
configuration (0.035-1.0~km baselines). This provided
sensitivity to emission on scales of up to 15$'$. The
primary beam of the VLA at our observing
frequency of $\sim$1420.3~MHz was $\sim 31'$. 

The VLA correlator was used in dual polarization (2AC) 
mode with a 0.78~MHz bandpass, yielding 256 spectral
channels with 3.05~kHz ($\sim$0.64~\kms) spacing. The band was centered
at a velocity of 25~\kms\ relative to the local standard of rest
(LSR); the band center was offset slightly from
the systemic velocity of the star ($V_{\rm sys,LSR}$=46.7~\kms)
to avoid placing a strong Galactic feature near the edge of the band.

Observations of Mira were interspersed with observations of two phase
calibrators (J0201-115 and J0220-019) approximately every 20
minutes. 3C48 was used as an absolute flux calibrator, and an
additional strong point source (J2253+161) was observed as a bandpass calibrator. To 
insure that the absolute flux scale and bandpass calibration were not
corrupted by Galactic emission in the band, the
flux and bandpass calibrators were each observed twice,
first with
the band shifted by $+$1~MHz and then by $-1$~MHz,
relative to the band center used
for the observations of Mira and the phase calibrators. We
estimate that this method yields an absolute flux scale accurate to
$\sim$10-15\%.

At the time of our
observations, the VLA contained 23 working antennas, 9 of which had
been retrofitted as part of the Expanded Very Large Array (EVLA) upgrade.
In total, 10.25 hours of integration were obtained on
Mira. However, some data were lost due to shadowing or hardware problems,
and significant
flagging to excise radio frequency interference (RFI) was
necessary, resulting in a loss of $\sim$13\% of the observed
visibilities. During our first observing session, roughly
half of the baselines had to be flagged in all channels numbering
integral multiples of 12, 13, and 14, owing to a strong local RFI
source that emitted an interference ``comb''. The source of this RFI
was identified as the Small Radio Telescope at the VLA Visitor Center, and
was switched off during the subsequent two observing sessions.

Our VLA data were calibrated and reduced
using the Astronomical Image Processing System
(AIPS). To avoid closure errors on VLA-EVLA baselines, we computed and
applied a bandpass solution to the raw data before proceeding with any
further calibration (G. van Moorsel, private communication).
A new frequency-averaged dataset was then
computed and used to calibrate the frequency-independent complex
gains (see Table~1). Following this, a second correction to the
bandpass was computed and applied, and time-dependent frequency
shifts were applied to the data to compensate for changes caused by
the Earth's motion. Finally,
prior to imaging, the $u$-$v$ data were
continuum-subtracted using a linear fit to the real and imaginary
components of the visibilities. Channels 20-85 and
105-160 were determined to be line-free and were used for these
fits. These channel ranges correspond to LSR velocities of
52.7$-$94.6~\kms\ and 4.4$-$39.8~\kms, respectively. The continuum
subtraction procedure was also effective at removing
frequency-independent patterns in the channel images caused
by solar contamination.

We imaged the VLA line data using the standard AIPS CLEAN deconvolution 
algorithm and produced data cubes using several different
weighting  schemes, two of which are presented here (Table~2). 
We also produced an image of the 21-cm continuum
emission in the region using a vector average of the line-free portion
of the
band. 

\clearpage
%
\begin{deluxetable}{lcccl}
\tabletypesize{\scriptsize}
\tablewidth{0pc}
\tablenum{1}
\tablecaption{VLA Calibration Sources}
\tablehead{
\colhead{Source} & \colhead{$\alpha$(J2000.0)} &
\colhead{$\delta$(J2000.0)} & \colhead{Flux Density (Jy)} & \colhead{Date}
}

\startdata
3C48$^{\rm a}$      & 01 37 41.2994 & +33 09 35.132 & 15.88$^{*}$   & All\\

0201-115$^{\rm b}$   & 02 01 57.1647 & $-$11 32 31.133 & 2.64$\pm$0.03 & 2007April1
\\
...       & ...           & ...           & 2.59$\pm$0.03 & 2007April30\\
...        & ...           & ...           & 2.65$\pm$0.04 & 2007May11\\

0220-019$^{\rm b}$  & 02 20 54.2800 & $-$01 56 51.800 & 3.33$\pm$0.06 & 2007April1
\\
...       & ...           & ...           & 3.33$\pm$0.10 & 2007April30\\
...        & ...           & ...           & 3.23$\pm$0.06 & 2007May11\\

2353+161$^{\rm c}$  & 22 53 57.7479 & +16 08 53.560 & 14.43$\pm$0.27$^{\dagger}$ &
2007April1\\
...      & ...           & ...           &14.29$\pm$0.29$^{\dagger}$ & 2007April30\\
...       & ...           & ...           &14.75$\pm$0.08$^{\dagger}$ & 2007May11\\

\enddata

\tablecomments{Units of right ascension are hours, minutes, and
seconds, and units of declination are degrees, arcminutes, and
arcseconds. }
\tablenotetext{*}{Adopted flux density at 1420.3~MHz,
computed according to the VLA
Calibration Manual (Perley \& Taylor 2003).}
\tablenotetext{\dagger}{Quoted flux density is the mean from the
two observed frequencies; see text.}

\tablenotetext{a}{Primary flux calibrator.}
\tablenotetext{b}{Secondary gain calibrator.}
\tablenotetext{c}{Bandpass calibrator.}

\end{deluxetable}


%
\begin{deluxetable}{lccccc}
\tabletypesize{\footnotesize}
\tablewidth{0pc}
\tablenum{2}
\tablecaption{Deconvolved Image Characteristics}
\tablehead{
\colhead{Image} & \colhead{{$\cal R$}} & \colhead{Taper} & 
\colhead{$\theta_{\rm FWHM}$} & \colhead{PA} & \colhead{rms} \\ 
\colhead{Descriptor} & \colhead{} & \colhead{(k$\lambda$,k$\lambda$)} & 
\colhead{(arcsec)} & \colhead{(degrees)} & \colhead{(mJy
beam$^{-1}$)} \\
\colhead{(1)} & \colhead{(2)} & \colhead{(3)} &
\colhead{(4)} & \colhead{(5)} & \colhead{(6)}  }

\startdata

Robust +1 & +1 & ... & $63''\times54''$ & $-3$ & 1.6-2.0\\

Tapered & +5 & 2,2 & $111''\times97''$ &+31 & 1.6-2.1 \\

Continuum & +1 & ... & $64''\times56''$ & $-10$ & 0.73\\


\enddata

\tablecomments{
Explanation of columns: (1) image or data cube designation 
used in the text; (2) robust
parameter used in image deconvolution (see Briggs 1995); 
(3) Gaussian taper applied in $u$ and
$v$ directions, expressed as
distance to 30\% point of Gaussian in units of kilolambda;
(4) dimensions of
synthesized beam; (5) position angle of synthesized beam (measured
east from north); (6) rms
noise per channel (1$\sigma$; line data) or in frequency-averaged
data (continuum).}

\end{deluxetable}

\clearpage

\section{VLA Results}
\subsection{The Morphology of Mira's \HI\ Envelope and Tail}
\begin{figure}

\scalebox{0.46}{\rotatebox{0}{\includegraphics{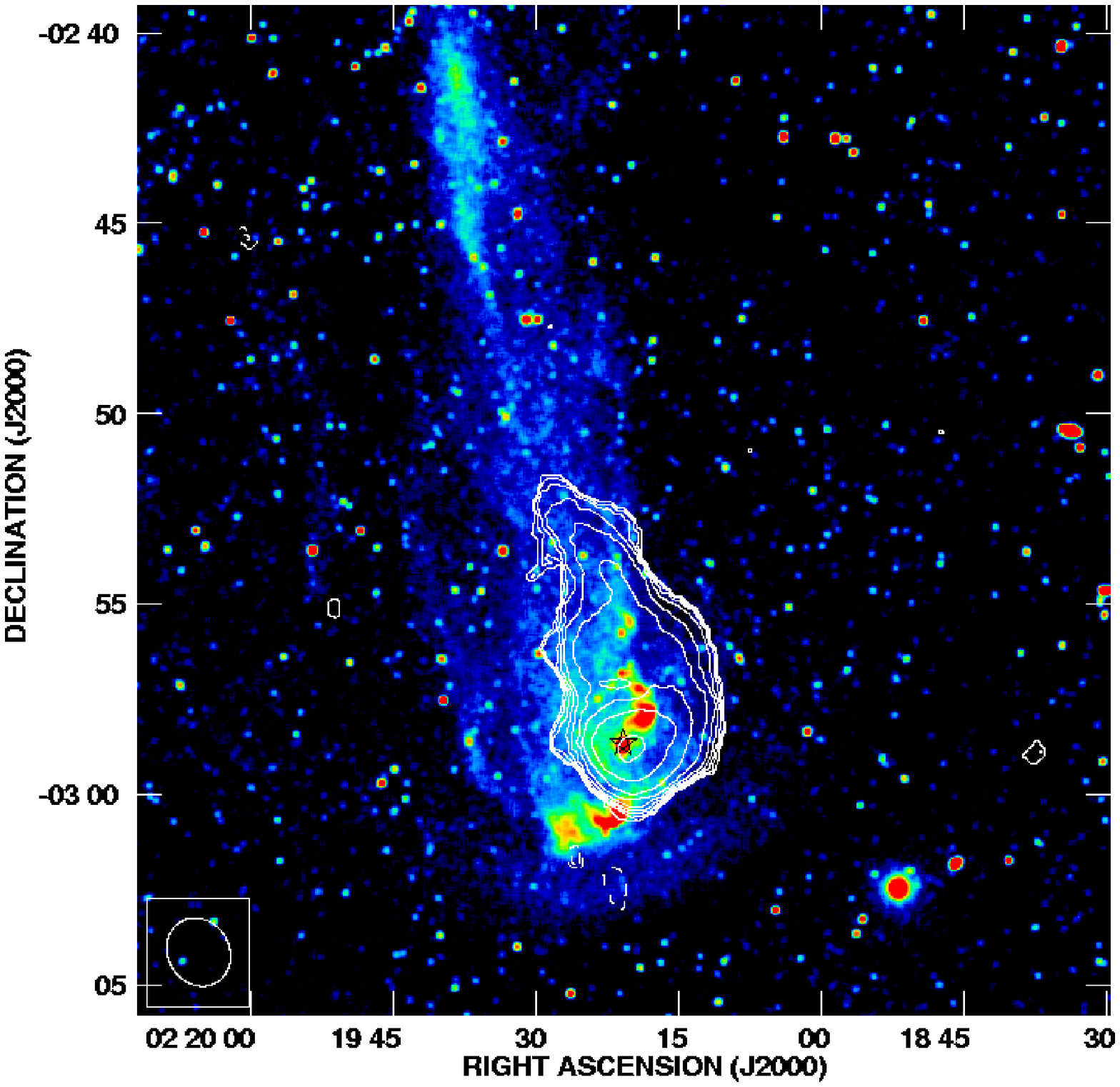}
\includegraphics{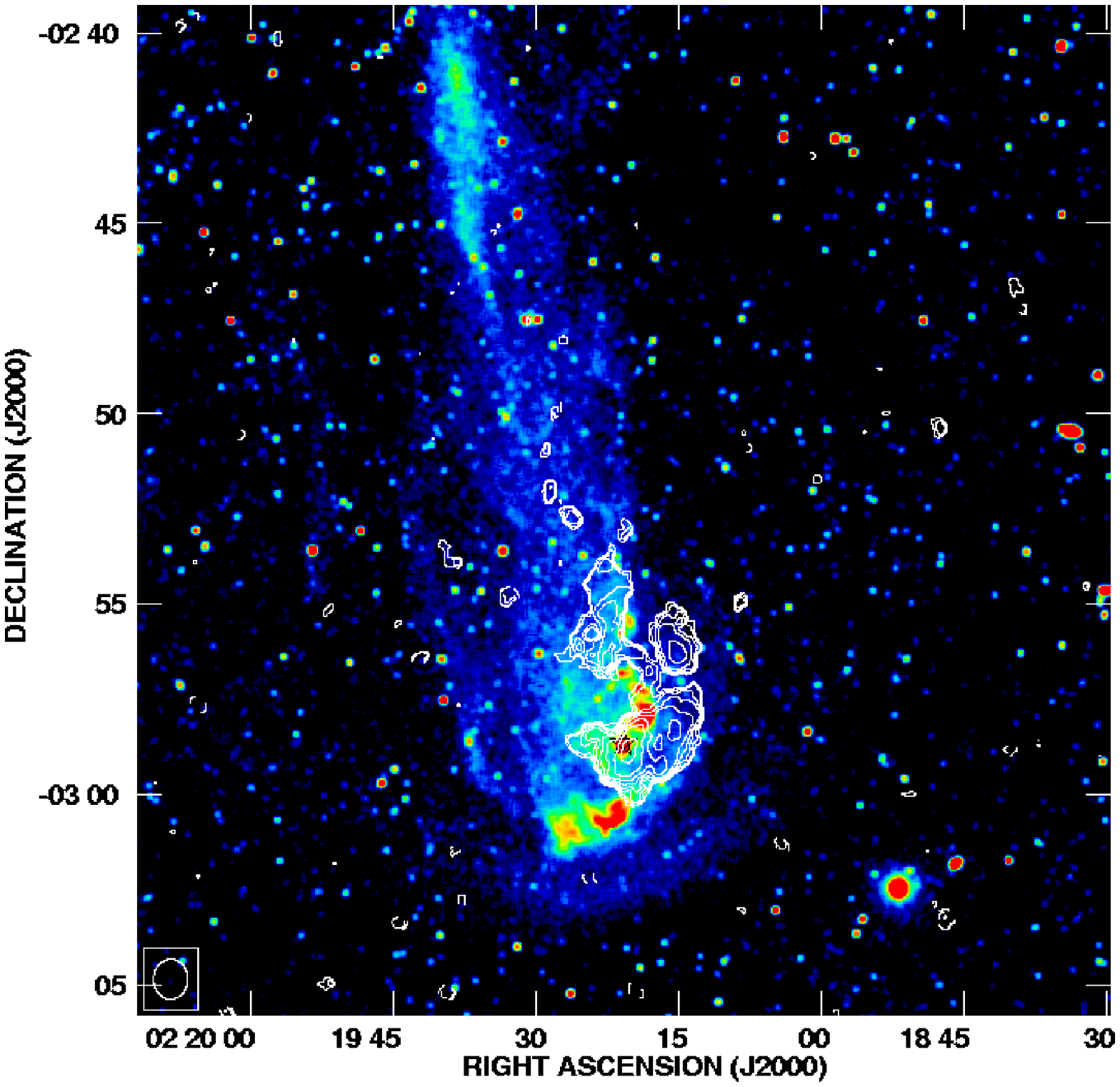}}}
\caption{\HI\ total intensity contours overlaid on false color 
{\it GALEX} FUV
images of Mira from M07. The {\it GALEX} image has been 
smoothed with a 3$\times$3 pixel (\as{4}{5}$\times$\as{4}{5})
boxcar function. The full extent of the FUV emission is not shown.
The left panel shows the \HI\ contours derived from the ``Tapered'' image
while the right panel shows those from the ``Robust +1'' image (see
Table~2). Contour levels are
($-2,1.4,-1,1,1.4,2,...22.4)\times3.5$ Jy 
beam$^{-1}$ m s$^{-1}$ (left); 
($-2,-1.4,1.4,2,...11.2)\times3.5$ Jy
beam$^{-1}$ m s$^{-1}$ (right). A black star symbol designates the 
position of Mira 
($\alpha_{J2000.0}$=02$^{\rm h}$~19$^{\rm m}$~20.79$^{\rm s}$, 
$\delta_{J2000.0}$=$-02^{\circ}$~$58'$~\as{39}{51}).
}
\protect\label{fig:mom0}
\end{figure}

Figure~\ref{fig:mom0} presents \HI\ total intensity contours for
Mira derived from our VLA imaging, 
overlaid on the {\it GALEX} FUV image from M07. \HI\ data with
velocities from $V_{\rm LSR}=40.5$ to $V_{\rm LSR}=50.1$~\kms\ were
included in these images. To improve
signal-to-noise in deriving the \HI\ maps, we rejected pixels in the
original data cubes
whose absolute values fell below 1.5$\sigma$ after
smoothing spatially with a Gaussian kernel of width 3 pixels
($30''$) and spectrally with a Hanning function. 

Our lower resolution \HI\ map (left) reveals 
a distinct ``head-tail'' structure, stretching
roughly 12$'$ ($\sim$0.4~pc) 
on the sky. 
[Note that the full extent of the \HI\ is significantly
greater than seen here (see \S~\ref{NRT}), but the VLA is insensitive to
the more extended emission.] 
We see that the brightest \HI\ emission
is concentrated near the position of
Mira itself. A trail of emission then  extends
to the northeast, following the same position angle as the FUV
tail. 

In our higher resolution \HI\ map (Figure~\ref{fig:mom0}, right), some
fraction of the total emission is lost (as it falls below our rejection
threshold), but 
we see that on smaller scales the \HI\ morphology of Mira 
becomes clumpy and
complex.  The location of the peak intensity of the \HI\ emission
shows a small but statistically significant offset 
to the southwest of the star's FUV position: ($\Delta\alpha,
\Delta\delta$)=($-$\as{12}{1}$\pm$\as{4}{3},$-$\as{14}{5}$\pm$\as{4}{8}).
This offset is comparable to the radius of the molecular envelope of
Mira found by Josselin et al. (2000)---consistent with the possibility
that the bulk of Mira's wind is molecular when it leaves the star, but
subsequently is partially dissociated, preferentially in the direction
of the leading edge of the shock front.
Close to the star, it is clear that 
the \HI\ emission is not symmetrically
distributed about Mira, but exhibits an enhancement
to the northwest. An enhancement in \KI\ emission was also seen along
this direction by Josselin et al. This type of asymmetry might
arise in part from
anisotropies in the outflowing wind and/or density gradients in the
surrounding ISM (see Vigelius et al. 2007). 
As the \HI\ emission  branches off to the north, it roughly follows
a ridge of bright FUV knots (part of what M07 term the ``North
Stream''), before bifurcating into two
lobes. A few additional isolated clumps of \HI\ are also visible 
to the north.

All of the \HI\ emission detected from Mira with the VLA 
overlaps with the FUV light
seen by {\it GALEX}, although the detailed relationship between the
two tracers is unclear.  \HI\ is seen
concentrated along the western side of the tail where the FUV emission
is also the brightest. 
However, a significant fraction of the FUV tail 
shows no \HI\ counterpart, including the bow shock,
the southeastern edge of the tail, and the FUV-bright region lying
between Mira and the bow shock (termed
the ``South Stream'' by M07). Moreover, on smaller scales 
there is no obvious correlation  
between the observed column density of
the \HI\ emission and the surface brightness of the FUV
emission. 
Detection of H$\alpha$ emission from the UV-bright knots by M07
suggests that most of the gas at these locations is likely to be
partially ionized. In the case of the South Stream, given that this region
has a different FUV$-$NUV color than the rest of the tail, the 
medium here may be very highly ionized, and that the FUV emission from this
location may have a different origin 
(e.g.,  hot plasma emission). 

\subsection{The Velocity Structure of the 
\HI\ Emission Surrounding Mira\protect\label{turbu}\label{vlaline}}
\begin{figure}
\centering
\scalebox{0.7}{\rotatebox{270}{\includegraphics{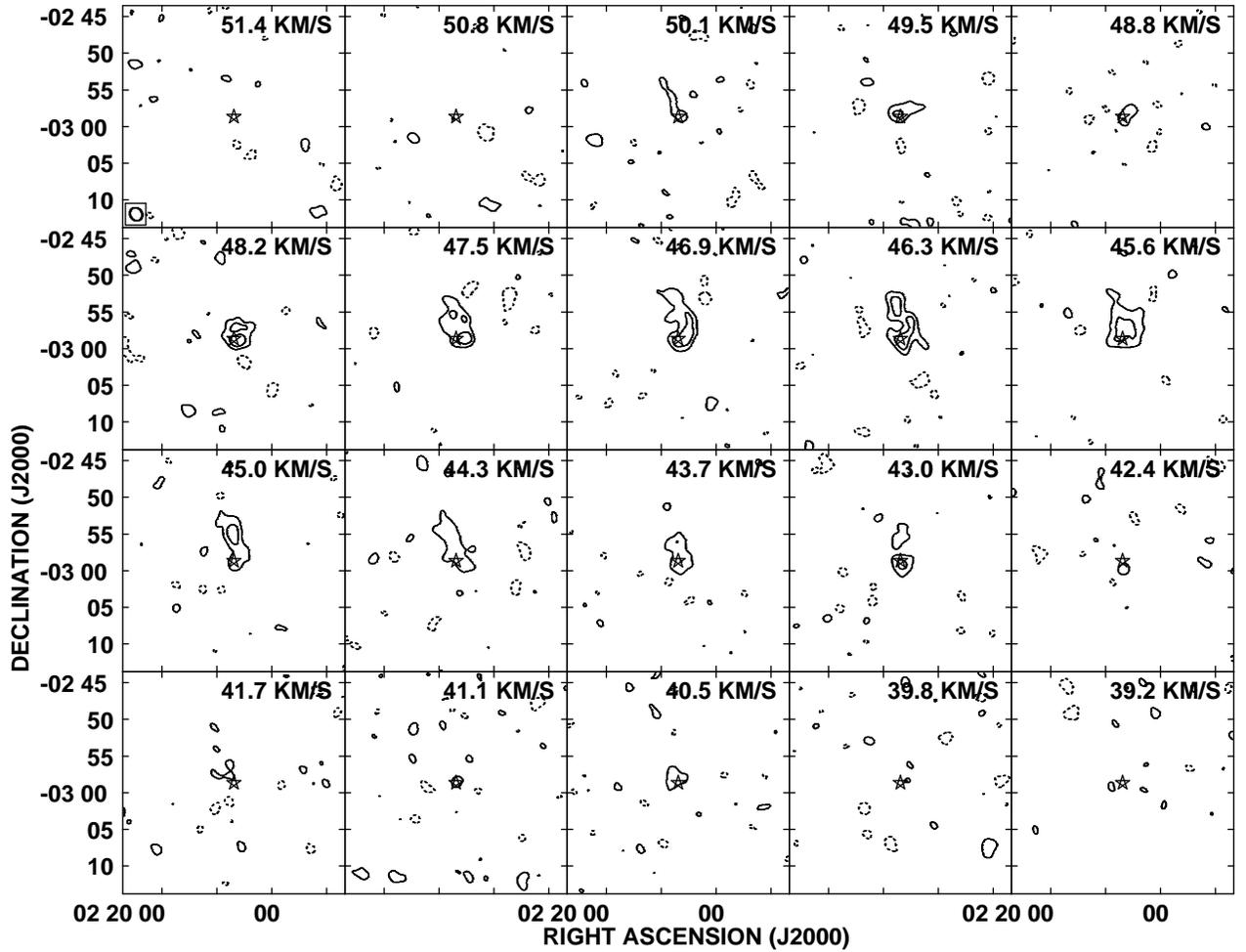}}}
\caption{\HI\ channel maps near the systemic velocity of Mira, taken
from the VLA ``Tapered'' data cube (Table~2). Contour levels are
($-5$[absent],$-2.5,2.5,5)\times1.8$ mJy beam$^{-1}$. A star symbol indicates the
position of Mira.}
\label{fig:cmaps}
\end{figure}

Individual \HI\ channel maps from our VLA imaging observations are shown in
Figure~\ref{fig:cmaps}. We find that near the position of the star, 
the emission detected in the central velocity maps
(44.3-47.5~\kms)
has a larger spatial extent than in the outer velocity
channels, as would be expected for an expanding envelope. At the same time,
several of the channels show additional 
emission extending toward the North that arises
from the near-tail. The velocity field of the latter component appears
complex, suggesting that the small-scale
motions of the tail gas may be affected by
turbulence. This is consistent with the interpretation of the tail as a
turbulent wake (e.g., Wareing et al. 2007b).

Figure~\ref{fig:totalspectra} shows the global \HI\ spectrum of Mira
derived from the VLA observations. The spectrum shown as a thick black
line was derived from the ``Tapered'' data cube (Table~2)
by summing all emission within a
$8'$(E-W)$\times$\am{13}{6}(N-S) box centered at $\alpha_{J2000.0}
=2^{\rm h}~19^{\rm m}~22.8^{\rm s}$,
$\delta_{J2000.0}=-2^{\circ}~54'~9''$.
Uncertainties on the total flux densities in each channel
are $\sim\pm$0.01~Jy. 
The VLA \HI\ profile agrees well with the NRT
line profile derived toward Mira and is
discussed further in \S~\ref{globsum}.

\subsection{Detection of \HI\ Absorption in the Tail\protect\label{absorb}}
\begin{figure}
\scalebox{0.60}{\rotatebox{90}{\includegraphics{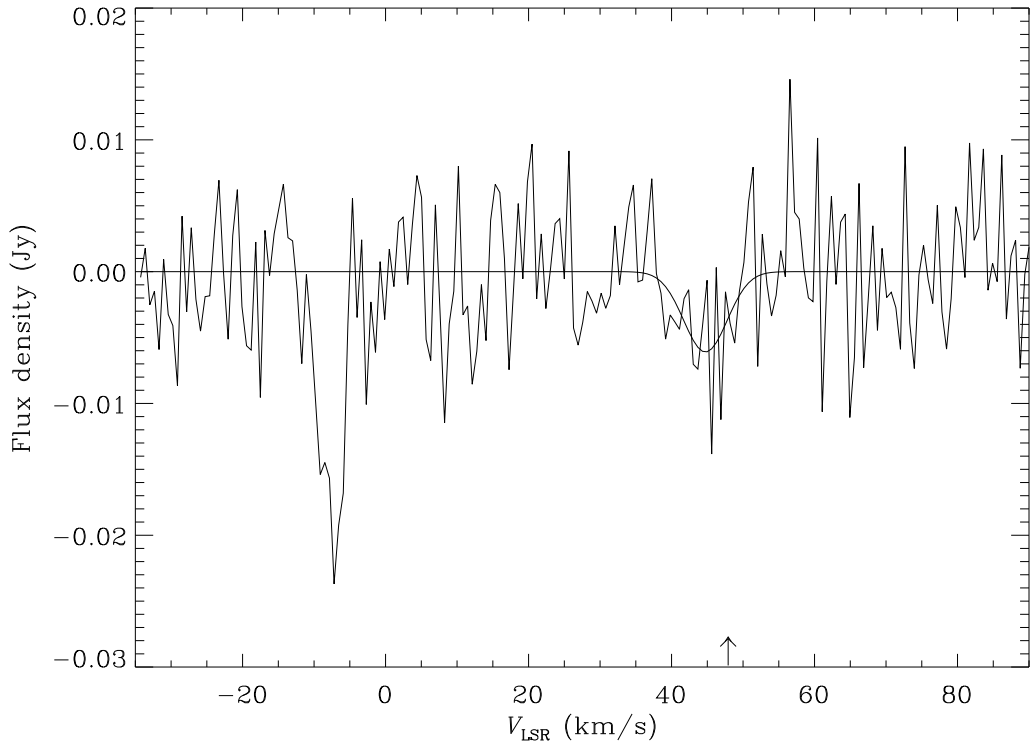}}}
\caption{\HI\ absorption spectrum toward the continuum source at 
$\alpha_{\rm J2000.0}=02^{h}19^{m}07.35^{s}$, 
$\delta_{\rm J2000.0}=-02^{\circ}52'$\as{49}{8}. The 
flux from the continuum source itself has been
subtracted. The stronger,  blueshifted absorption
feature near $-8$~\kms\ is due to Galactic interstellar material along the
line-of-sight, but the weaker, 
redshifted feature has a velocity and linewidth consistent with 
the circumstellar material surrounding Mira. 
The thick line shows a Gaussian
fit to the latter feature (see text for details). An arrow indicates the
stellar systemic velocity of Mira determined from CO observations. }
\label{fig:Mira_absorb}
\end{figure}

\begin{figure}
\centering
\scalebox{0.7}{\rotatebox{-90}{\includegraphics{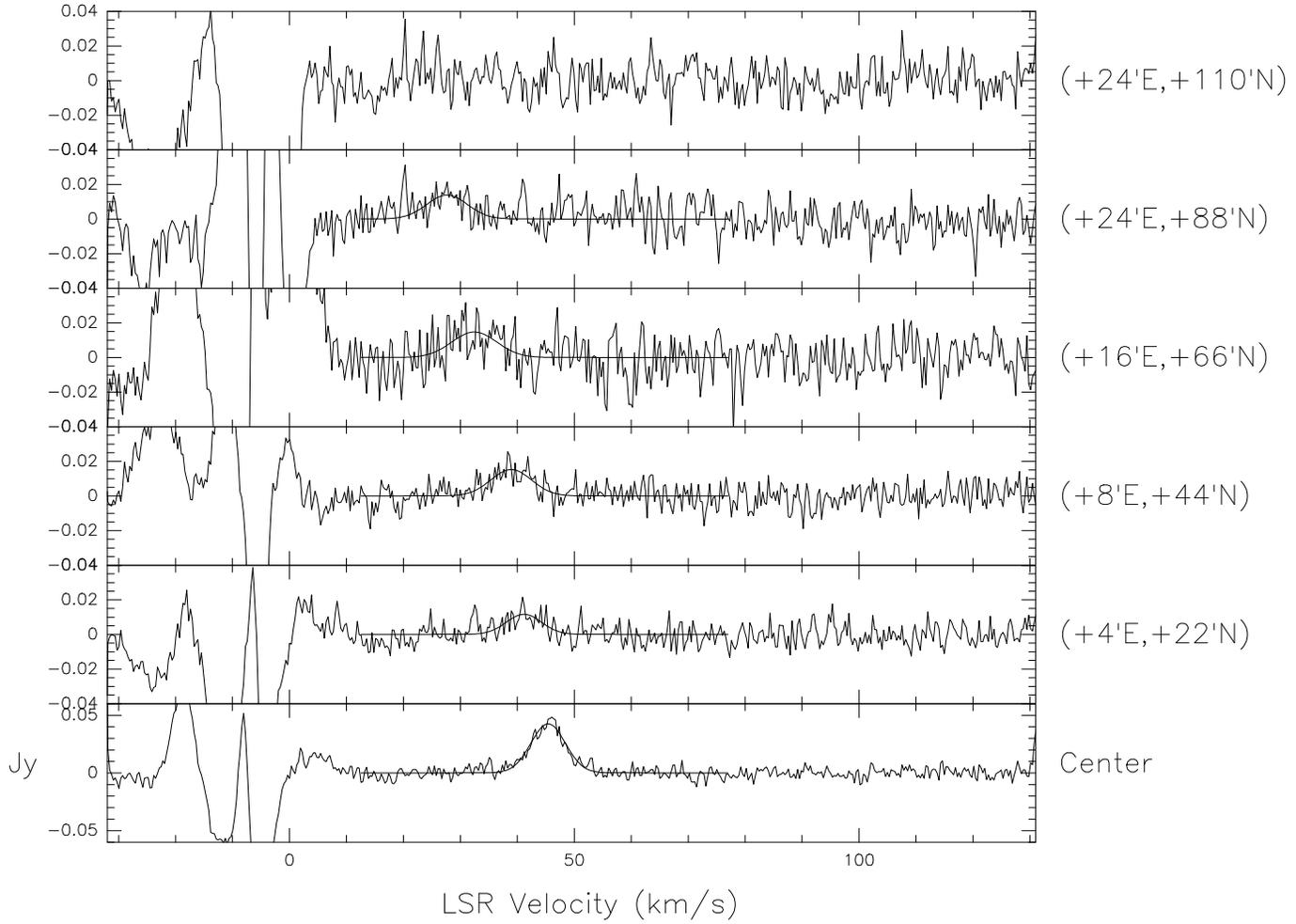}}}
\caption{NRT \HI\ spectra along Mira's tail. The spectra shown have
been smoothed to a velocity resolution 
of 0.32~\kms. Gaussian fits to the emission from Mira are
overplotted. Note the 
bottom panel has a different vertical scale. Features blueward of
$V_{\rm LSR}\approx$10~\kms\ are from intervening Galactic emission.}
\label{fig:nancay}
\end{figure}

The 21-cm continuum emission within a 30$'$ region surrounding Mira comprises
a number of weak point sources
with a total observed flux density of $\sim$0.4~Jy (uncorrected for
primary beam attenuation). Mira~AB itself is
undetected, and we detect no continuum emission coincident with any of
the bright knots seen in the UV and in H$\alpha$ by M07.
The brightest continuum source in the region lies at
$\alpha_{\rm J2000.0}=02^{\rm h}~19^{\rm m}~07.36^{\rm s}$, 
$\delta_{\rm J2000.0}=-02^{\circ}~52'$~\as{49}{2}, and we measure for it 
a flux density of 0.228$\pm$0.003~Jy (after correction 
for the primary beam).  It overlaps 
with the FUV emission detected by {\it GALEX}, but 
lies outside the region where \HI\ was detected in emission
with the VLA. We have examined a spectrum toward
this source and detect a weak
($\sim$3$\sigma$) absorption feature
(Figure~\ref{fig:Mira_absorb}). Based on a Gaussian fit, this feature has
a peak flux density $S_{0}=-6.1\pm1.9$~mJy, a FWHM of $\Delta
v=7.1\pm$0.9~\kms, 
and a central velocity 
$V_{\rm LSR}=44.7\pm$0.9~\kms. 
Both the central velocity and the 
width of the line feature are consistent with  
the \HI\ gas observed in the tail of Mira in 
emission (see Figures~\ref{fig:totalspectra} and \ref{fig:nancay}). 

Detection of \HI\ in
absorption in the tail of Mira 
allows us to obtain a constraint on the spin temperature
of the gas. For the ``Robust +1'' data, 
the limiting \HI\ column density for
a detection of \HI\ in {\it emission}, integrated over a  Gaussian
line profile with FWHM 7.1~\kms, is
$N_{\rm HI}\lsim1.3\times10^{19}$~cm$^{-2}$ (3$\sigma$). 
Under the assumption that 
the absorbing gas at the position of the continuum source 
has an equal or lower column density than gas detected in emission, 
one may then write:
\begin{equation}
T_{s}\le \frac{N_{\rm HI}}{(1.8\times10^{18})\int\tau(v)dv} ~{\rm K}
\end{equation}
\noindent where $T_{s}$ is the spin temperature of the atomic hydrogen and
$\tau$ is its optical depth (e.g.,
Dickey et al. 1978). The assumption of a
Gaussian line shape yields a line-integrated optical depth for the \HI\
absorption profile of
$\approx$0.20, and therefore
$T_{s}\lsim$37~K. Clumping of the absorbing  material would further
reduce this limit, implying that a component of the tail gas
is rather cool. 

\section{Observations with the Nan\c{c}ay Radio Telescope\protect\label{NRT}}
To achieve greater sensitivity to extended, low surface brightness
\HI\ emission in the vicinity of Mira,
we made  observations at several positions along its tail
with the NRT (see Table~3). 
Our pointings were selected using the {\it GALEX} map from M07 as a
guide. 
These observations were obtained between 2007 
September and 2007 December 
as part of a Target of Opportunity program. 

The NRT is a meridian-transit-type telescope with an effective
collecting area of roughly 4000~m$^{2}$.
At 1420 MHz, its half-power beam width is $4'$ in right 
ascension and 22$'$ in declination for a source at the declination of
Mira ($-2^{\circ}$). Typical system temperatures are
$\sim$35~K. Further properties of the NRT are described in van Driel
et al. (1997). The large collecting area of the NRT 
and the good match between the N-S extension of the beam
and the direction of Mira's wake make the NRT well-suited to 
searching for extended, low column density material. 

Our observational strategy for mapping the tail consisted of
position-switched measurements at each pointing, with beam-throws 
of $\pm12'$ or $\pm16'$ in the E-W direction. One-third of the time 
was devoted to the on-position and two-thirds of the time to the 
off-source comparison spectra. 
A full NRT spectrum has a bandwidth of 165~\kms\ and a spectral 
resolution of 
0.08~\kms. 
A total of 44 hours of data were obtained along the tail. 
Fortunately, there is minimal Galactic \HI\ emission near the LSR
velocity of Mira; this provides flat baselines that permit us to detect
weak signals efficiently. Data processing was performed using the CLASS
software and consisted of subtracting a linear baseline from each
spectrum before averaging. 

The results of our NRT mapping are summarized in Table~3, and 
we show a sampling of our spectra in
Figure~\ref{fig:nancay}. We have clearly detected \HI\ emission from
Mira's tail as far as $88'$ north of the star. Moreover, we see the peak
velocity of the emission becomes progressively blueshifted with increasing
distance from the star, indicating an overall deceleration (see also 
\S~\ref{age}).
We did not detect \HI\ in any of the NRT pointings that have no overlap
with Mira's FUV tail, consistent with the
material giving rise to the FUV light 
and the \HI\ being spatially
coupled along the full length of the tail. 

\clearpage
%
\begin{deluxetable}{lccccc}
\tabletypesize{\footnotesize}
\tablewidth{0pc}
\tablenum{3}
\tablecaption{NRT Mapping of Mira's Tail}
\tablehead{
\colhead{Position offset:$^{\rm a}$} & \colhead{Integration
time} & \colhead{rms noise} & \colhead{Velocity} &
\colhead{Line width} & \colhead{$F_{\rm peak}$} \\
\colhead{(arcmin E, arcmin N)} & \colhead{(hours)} & \colhead{(mJy)} & 
\colhead{(\kms)} & \colhead{(\kms)} & \colhead{(mJy)} 
}

\startdata

(0,0)     & 49 &  4.56  & 45.4$\pm$0.5 & 7.1$\pm$0.3 & 42.51$\pm$0.46 \\
(0,+22) & 4 & 8.85 & ... & ... & ... \\
(+4,+22)  & 7  &  5.78  & 41.2$\pm$0.5 & 7.3$\pm$1.2 & 11.6$\pm$1.7\\
(+8,+22) & 2 & 10.30 & ... & ... & ... \\
(+8,+44)  & 7  &  6.42  & 38.9$\pm$0.5 & 8.4$\pm$1.1 & 15.2$\pm$1.7\\
(+12,+66) & 4 & 7.64 & ... & ... & ... \\
(+16,+66) & 4  &  11.10 & 32.4$\pm$0.9 & 9.3$\pm$2.1 & 14.7$\pm$2.9\\
(+18,+88) & 4 & 11.30 & ... & ... & ...  \\
(+20,+88) & 4 & 9.04 & ... & ... & ...  \\
(+24,+88)& 4  & 9.19 & 27.7$\pm$0.7 & 8.4$\pm$1.6 & 13.9$\pm$2.3\\
(+24,+110) & 4 & 9.11 & ... & ... & ... \\

\enddata

\tablecomments{Tabulated line parameters are derived from Gaussian
fits to the spectra shown in Figure~\ref{fig:nancay}. }

\tablenotetext{a}{The adopted 
coordinates of Mira were 
$\alpha_{J2000.0}$=02$^{\rm h}$~19$^{\rm m}$~20.79$^{\rm s}$, 
$\delta_{J2000.0}$=$-02^{\circ}$~$58'$~\as{39}{51}.}

\end{deluxetable}

\clearpage

\section{Results and Interpretation\protect\label{results}}

\subsection{The Global \HI\ Line Profile and Total \HI\ Mass of 
Mira's Circumstellar Material\protect\label{globsum}}
\begin{figure}
\centering
\scalebox{0.7}{\rotatebox{90}{\includegraphics{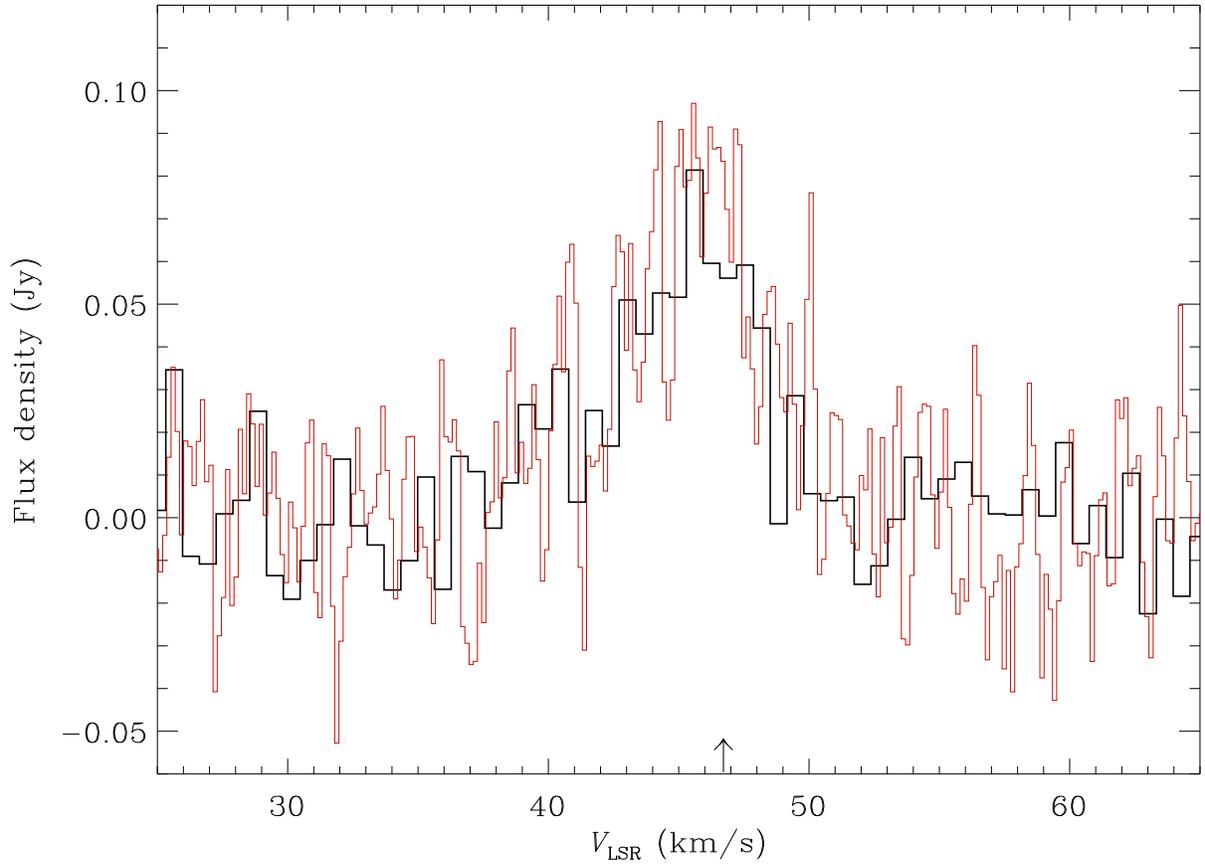}}}
\caption{\HI\ spectra toward Mira. The thin red line shows the NRT spectrum
obtained by summing the measurements over a 
12(E-W)$'\times22'$(N-S) region; the thick black line shows the VLA spectrum obtained
by summing within a \am{8}{0}$\times$\am{13}{6} region. The arrow
indicates the stellar systemic velocity obtained from CO observations
by Winters et al. 2003.}
\label{fig:totalspectra}
\end{figure}

The region near the position of 
Mira has been extensively observed with the 
NRT since 2000
(see also G\'erard \& Le~Bertre 2006). The \HI\ line spectrum 
we derive by integrating the emission throughout a 12(E-W)$'\times22'$(N-S) region
agrees well with the integrated line profile obtained with the VLA
(Figure~\ref{fig:totalspectra}). Based on
Gaussian fits to the global line profiles from the two telescopes, we
find line centroids of 45.69$\pm$0.20~\kms\ and 45.41$\pm$0.26~\kms\
for the NRT and VLA, respectively. These centroids are
slightly blueshifted compared with the value derived from 
CO(2-1) line observations by Winters et al. (2003; $V_{\rm
CO}=46.7\pm0.3$~\kms). We note however that the CO line is somewhat
asymmetric and appears to be comprised of multiple 
components. Based on a two-component fit to the CO(3-2) spectrum, Knapp et
al. (1998) find the broader component to be slightly
blueshifted ($V_{\rm CO}=46.0\pm$1.0~\kms), making it consistent 
with the \HI\ centroid to within uncertainties.
The FWHM of the \HI\ profiles are 6.64$\pm$0.20~\kms\ (NRT) and
6.13$\pm$0.26~\kms\ (VLA), comparable to linewidths measured from CO
data (Knapp et al. 1998; Winters et al. 2003). However, whereas the CO linewidths
directly gauge the expansion velocity of the stellar wind, the \HI\ profile
width may be affected by turbulent motions in the tail gas
(\S~\ref{turbu}) or by possible thermal 
broadening (Libert et al. 2007).

Integrating over the line profiles shown in Figure~\ref{fig:totalspectra}
yields integrated flux densities of 0.47$\pm$0.04~\jks\
(VLA) and
0.51$\pm0.03$~\jks\ (NRT). Assuming the \HI\ is optically thin, 
the total \HI\
mass contained within the portion of the circumstellar material 
imaged by the VLA is
$M_{\rm HI}\approx1.27(\pm 0.13)\times10^{-3}~M_{\odot}$. 
Using the NRT measurements summarized in Table~3, we can also estimate
the additional amount of atomic material in the extended tail 
to be $M_{\rm HI}\sim2.7\times10^{-3}~M_{\odot}$. (Here we
have multiplied the observed emission by a geometric
correction factor of 2 to account for the fact that we have not fully
sampled the tail). The combined \HI\
mass for the circumstellar envelope and tail of Mira is then $M_{\rm
HI}\sim4\times10^{-3}~M_{\odot}$.

\subsection{A Revised Age for Mira's Tail\protect\label{age}}
A key result of our NRT mapping is that the spectra reveal 
a clear slowing-down of the material in the tail with
increasing distance from Mira (Table~3 and Figure~\ref{fig:nancay}). 
At $\sim$\ad{1}{5} from the star, the peak \HI\ signal is 
$\sim$14~mJy, with $V_{\rm LSR}=27.7\pm0.7$~\kms, 
whereas the centroid of the \HI\ signal at the center
position is 45.4$\pm$0.5~\kms. This finding is consistent with the model 
of Wareing et al. (2007b), who predicted an increasing 
velocity lag 
with respect to the velocity of Mira itself with increasing distance
along the tail. This result also implies
that the tail is older than the age  of 3$\times10^{4}$~years derived by 
M07 under the assumption that this material
is stationary with respect to the ISM.

From Table~3, we can extrapolate to estimate 
the radial velocity of the \HI\ material
at 2$^{\circ}$ from Mira (i.e., at the most extreme position
where $GALEX$ detected emission) to be $V_{\rm LSR}\sim23\pm3$~\kms. 
Adopting the
stellar radial velocity determined from CO observations
($V_{\rm LSR}=$46.7~\kms; 
Winters et al. 2003), the proper motion from Perryman et al. (1997), and
the solar motion parameters from Dehnen \& Binney (1998),
we
estimate Mira's velocity in the plane of the sky, corrected
for solar motion,
to be $V_{t}\approx$120~\kms. The velocity lag for the outermost tail
material is therefore 23$\pm$3~\kms\ in the radial direction
and $\sim61.0\pm$7.7~\kms\ in the plane of the sky. 
Finally, assuming a uniform deceleration of the stellar gas, we derive
an age of $t\sim (1.20^{+0.17}_{-0.14})\times10^{5}$ years 
for the material detected by
$GALEX$ at 2$^{\circ}$ from Mira. This calculation does not 
take into account a
possible variation of the mass-loss rate or the turbulence of the
interstellar medium that the stellar gas may encounter.

Our age estimate for Mira's tail exceeds previous observational
estimates for Mira's total mass-loss duration by roughly an order of
magnitude or more (see Young et al. 1993; Bowers \& Knapp 1988;
G\'erard \& Le~Bertre 2006; M07).
Furthermore, the tail age approaches the expected interval between two thermal
pulses; the relatively modest change in surface brightness over the
length of Mira's tail then suggests that the predicted
growth in the mass-loss rate for AGB stars between thermal pulses 
(e.g., Vassiliadis \& Wood 1993) may be smaller than previously assumed.

\subsection{Comments on the Composition of Mira's Circumstellar
Envelope and Tail}
Adopting the mass-loss rate for Mira derived from CO observations (${\dot
M}\sim1.7\times10^{-7}~M_{\odot}$ yr$^{-1}$; Ryde \& Sch\"oier 2001) 
and assuming this mass-loss
rate has remained constant in time, the age derived in \S~\ref{age} implies
that the total mass of Mira's
circumstellar debris should be $\sim2.0\times10^{-2}~M_{\odot}$. After
adjusting our current \HI\ measurements 
for the mass of He, we then estimate that neutral atomic
material accounts for 
$\sim$25\% of Mira's circumstellar envelope and tail. We now 
briefly comment 
on some possible implications of this finding.

Previous observations have shown
that Mira's wind is likely to be predominantly molecular as it leaves
the star (e.g., Bowers \& Knapp 1988; Josselin et al. 2000; Wood et al. 2002). However, 
as discussed by Josselin et al., the bulk of Mira's
circumstellar material is expected to be dissociated by the
interstellar radiation field at radii of $r\gsim2\times10^{16}$~cm
from the star.  
Therefore, unless the wind is very
clumpy (thereby increasing the survival time of the molecules), it is
expected that atomic matter will comprise a significant
fraction of the material that is ultimately swept by ram pressure into the tail.

Under the assumption that the FUV
light from Mira's tail arises entirely
from collisional excitation of H$_{2}$ by hot
electrons (M07), an expected by-product will be rapid
dissociation of molecules (see also Raymond et al. 1997), thus
providing an additional atomic contribution to Mira's tail. 
Indeed, the
dissociation rate of $\sim2.5\times10^{42}$~s$^{-1}$ 
assumed by M07 should have produced roughly a factor of four 
more H atoms during the past $1.2\times10^{5}$~years
than we observe. 
Assuming some fraction of
the wind is atomic before being swept into the tail, this raises some
difficulty in how to maintain a sufficient supply of H$_{2}$  to power
Mira's FUV luminosity 
over its inferred lifetime. A significantly lower molecular
dissociation rate ($\sim$10\%) could help resolve this problem;
however such a rate approaches that expected from the interstellar
radiation field alone (Morris \& Jura 1983). If a portion of the material in
Mira's tail is clumpy, self-shielding of the molecular hydrogen could
also help to increase its lifetime (see, e.g., Huggins et
al. 2002). A search for CO emission associated with
such a clumped component in Mira's tail would be of considerable interest. 
An alternative explanation may be that some
fraction of the FUV light from Mira's tail arises from processes
involving atomic material, such as the H two-photon continuum (e.g.,
D'Odorico et al. 1980)
and/or bremsstrahlung emission from a hot ($T\sim10^{5}$~K) 
``surface'' of the tail.
Future multi-wavelength observations and 
modelling should help to clarify these issues.

\section{Discussion: Are \HI\ Tails Ubiquitous Features of Evolved Stars
Undergoing Mass-Loss?\protect\label{discussion}}
We have reported the detection of an extended tail of neutral, atomic
hydrogen 
associated with the AGB star Mira. This \HI\ 21-cm line emission
arising from the tail coincides with the
FUV-luminous wake recently discovered by M07. Although Mira is
currently the only known star to have a FUV-bright tail, we draw
attention to the possibility that 
its \HI\ tail may represent an extreme example of a 
rather common
phenomenon for evolved stars undergoing mass-loss.

G\'erard \& Le~Bertre (2006) already reported evidence that \HI\
emission associated with circumstellar envelopes may be offset from
the position of the central star. In addition,
Matthews \& Reid (2007) previously reported the detection of
an \HI\ ``plume'' stretching $\sim$0.2~pc from the semi-regular
variable star RS~Cnc. We have since confirmed that the geometry 
of this plume is consistent with material trailing the motion of RS~Cnc
through the ISM. More recently, we have imaged an analogous,
but somewhat shorter tail associated with another semi-regular
variable star, X~Her (Gardan et al. 2006; Matthews et al., in prep.).  
Both of these stars have
smaller space velocities than Mira ($\sim$18~\kms\ and $\sim$100~\kms, 
respectively), indicating that unusually high space motion is not a
prerequisite for tail formation; indeed, it may require only that the
stellar space velocity exceeds the expansion velocity of the wind.

While the sample of stars
imaged in \HI\ is presently small, evidence of interaction
between the circumstellar envelope and the ISM has also been seen in the
global \HI\ spectra of a number of \HI-detected stars (e.g., G\'erard \&
Le~Bertre 2006 and references therein). Observed \HI\ line
profile shapes are frequently 
inconsistent with a classic spherically symmetric
model of mass-loss at a constant outflow speed, and may show velocity
centroids offset from those observed in CO. As shown by
Gardan et al. (2006) and Libert et al. (2007), these profiles
can be well
reproduced once the effects of ISM interaction are accounted for. The
importance of ISM interactions in the evolution of
circumstellar envelopes has also been underscored by the numerical
simulations of Villaver et al. (2002) and 
Wareing et al. (2007a,c), and by the discovery of a far-infrared bow
shock associated with the AGB star R~Hya (Ueta et
al. 2006). {\it We therefore
propose that extended gaseous tails may be ubiquitous
features of evolved stars undergoing mass-loss.} 
For stars with low space velocities, hot
companions\footnote[6]{Mira's hot companion, Mira~B, is
unlikely to significantly affect the composition of Mira's wind and
tail owing
to the small extent of the ionized zone surrounding it (see Matthews
\& Karovska 2006).}, and/or largely atomic winds, these
tails may lack associated bow shock structures and/or 
a detectable FUV counterpart, but should in many instances 
be readily
detectable via \HI\  21-cm line observations.
The conditions for detection of these tails 
will be most favorable for stars at high
Galactic latitudes and/or with systemic  velocities well removed from
those of the bulk of the Galactic emission.

\acknowledgements

We thank R. Perley and the VLA staff for tracking down the RFI problem
affecting our early observations. We are also grateful to M. Seibert
for providing us with the {\it GALEX} FUV image and to
J. Raymond for valuable discussions. 
The \nan\ Radio Observatory is the Unit\'e Scientifique \nan\ 
of the Observatoire de Paris and is associated with the 
French Centre National de Recherche Scientifique (CNRS) as the Unit\'e
de Service et de Recherche (USR), No. B704. 
The Observatory also gratefully acknowledges the financial support of
the R\'egion Centre in France. 
The VLA observations presented here were part of program AM887.

\clearpage

\clearpage

\clearpage

\clearpage

\clearpage

\end{document}